\documentclass{PoS}
\newcommand{\bea}{\begin{eqnarray}}
\newcommand{\eea}{\end{eqnarray}}
\usepackage{amsmath}
\usepackage{amsfonts}
\usepackage{amssymb}
\newcommand{\nn}{\nonumber}

\title{Impact of leptoquarks in semileptonic $B$ decays}

\ShortTitle{Impact of leptoquarks in semileptonic $B$ decays}

\author{\speaker{Suchismita Sahoo}
\thanks{The authors  would like to thank Science and Engineering Research Board (SERB),
Government of India for financial support through grant No. SB/S2/HEP-017/2013.}\\
        School of Physics, University of Hyderabad,\\ Hyderabad-500046, India\\
        E-mail: \email{suchismita@uohyd.ac.in}}

\author{Rukmani Mohanta\\
        School of Physics, University of Hyderabad,\\ Hyderabad-500046, India\\
        E-mail: \email{rmsp@uohyd.ernet.in}}
 
\author{Anjan K. Giri\\
        Physics Department, IIT Hyderabad,\\ Kandi - 502285, India\\
        E-mail: \email{giria@iith.ac.in}}        

\abstract{We study the impact of  vector leptoquarks on  the recent anomalies in  semileptonic $B$ meson decays such as $R_K$ and $R_{D^{(*)}}$ etc. We constrain the leptoquark couplings by using the existing data on the branching ratios of $B_s \to l^+ l^-$ and $\bar{B} \to X_s l^+ l^-$  processes, where $l=e,\mu,\tau$. We estimate the branching ratios of $\bar B \to D^{(*)} l \bar \nu_l$ processes using the constrained leptoquark parameter space. We also investigate the possibility of  simultaneous  explanation  of  $R_{D^{(*)}}$, $R_K$ anomalies in this model.}

\FullConference{9th International Workshop on the CKM Unitarity Triangle\\
		28  November - 3 December 2016\\
		Tata Institute for Fundamental Research (TIFR), Mumbai, India}

\begin{document}

\section{Introduction}
Recently both BaBar \cite{BaBar} and Belle \cite{Belle} experiments have measured the  ratio of branching fractions of $\bar B \to  D \tau \bar \nu_\tau$  over $\bar B \to  D l \bar \nu_l$ decays, where $l=e$, $\mu$  and the present average experimental result \cite{hfag} is  
\bea
R_{D}=\frac{{\rm Br}\left(\bar{B} \to D \tau \bar \nu_\tau \right)}{{\rm Br}\left(\bar{B} \to D l \bar \nu_l \right)} = 0.397 \pm 0.040 \pm 0.028, 
\eea
which has  $1.9\sigma$ deviation from its standard model (SM) prediction  
$R_D^{\rm SM} = 0.300 \pm 0.008$ \cite{RD-SM}. Besides, both the $B$ factories and LHCb \cite{LHCb} have reported  $3.3\sigma$ discrepancy \cite{hfag}  in the measurement of $R_{D^*}$ 
\bea
R_{D^*}=\frac{{\rm Br}\left(\bar{B} \to D^* \tau \bar \nu_\tau \right)}{{\rm Br}\left(\bar{B} \to D^* l \bar \nu_l \right)} = 0.316 \pm 0.016 \pm 0.010,
\eea
from its SM result $R_{D^*}^{\rm SM} = 0.252 \pm 0.003$ \cite{RD*-SM}.  Analogously another interesting observable  is the lepton nonuniversality (LNU) parameter $(R_K)$ in  $B^+ \to K^+ l^+ l^-$ process,  which has recently been measured at   LHCb,  with value  \cite{LHCb-RK},
\bea
R_K=\frac{{\rm Br}\left(B^+ \to K^+ \mu^+ \mu^- \right)}{{\rm Br}\left(B^+ \to K^+  e^+ e^- \right)} = 0.745^{+0.090}_{-0.074} \pm 0.036,
\eea
and corresponds to a $2.6\sigma$ deviation from its SM   value $R_K = 1.0003 \pm 0.0001$ \cite{Hiller} in the dilepton invariant mass squared bin $\left( 1 \leq q^2 \leq 6 \right) {\rm GeV^2}$.

In this paper, we  would like to investigate the  semileptonic $\bar B \to D^{(*)} l \bar \nu_l$  decay processes, mediated by the FCNC transitions $b \to c l \bar \nu_l$ in the vector leptoquark (LQ) model. We compute the branching ratios of $\bar B \to D^{(*)} l \bar \nu_l$  modes in this  model. The simultaneous  study of   the lepton non-universality  parameters $R_{D^{(*)}}$ and $R_{K^{(*)}}$ is the main objective of this work.   LQs are  color triplet hypothetical bosonic particles which allow quarks and leptons to interact simultaneously and carry both baryon and lepton numbers. LQs can have spin 0 (scalar) or spin 1 (vector) and  are encountered in various extensions of the SM, such as technicolor model, GUT theories, Pati-Salam models and the quark and lepton composite model. 

The outline of the paper is follows. In section 2, we discuss the effective Hamiltonians involving $b \to c l \bar \nu_l$ and $b \to s ll$ transitions in the SM. We also present the new physics contribution due to the vector LQ  exchange. The constraint on LQ parameter space by using the experimental limit on the branching ratios of the $B_s \to l^+ l^-$ and $\bar B \to X_s l^+ l^-$ decays and the numerical analysis for $\bar B \to D^{(*)}l \bar \nu_l$ processes are presented  in section 3.  Section 4 contains the conclusion.
\section{Effective Hamiltonian and the  new physics contribution from vector leptoquark exchange}
In the SM, the effective Hamiltonian describing the processes mediated by the $b \to c l \bar  \nu_l$ transition is given by \cite{sakaki} 
\bea \label{ham-bc}
\mathcal{H}_{eff}=\frac{4G_F}{\sqrt{2}} V_{cb} \Big [ \left(\delta_{l\tau} + C_{V_1}^l \right) \left(\bar{c}_L \gamma^\mu b_L \right) \left(\bar{\tau}_L \gamma_\mu \nu_{lL} \right) \Big ],
\eea
where $G_F$ is the Fermi constant, $V_{cb}$ is the CKM matrix element and $q_{L(R)} = L(R)q$ are the chiral quark fields with $L(R) = (1 \mp \gamma_5 )/2$ as the projection
operators. The Wilson coefficient $C_{V_1}^l$ is zero in the SM and can only be generated by the new physics model. 

The effective Hamiltonian for $b \to s l^+ l^-$ process in the SM is given by
\bea
{\cal H}_{eff} &=& - \frac{ 4 G_F}{\sqrt 2} V_{tb} V_{ts}^* \sum_{i=1}^{10} C_i(\mu) \mathcal{O}_i (\mu)\;,\label{ham-bs}
\eea
where $\mathcal{O}_i$'s  are the six dimensional operators and  $C_{i}$'s are the corresponding  Wilson coefficients.

Models with vector  LQs can modify the SM effective Hamiltonian (\ref{ham-bc}, \ref{ham-bs}) due to the additional contributions arising from the   LQs exchange and will give measurable deviations from the predictions of the SM in the beauty sector. Here we consider $U_3(3,3,2/3)$ vector LQ multiplet which is invariant under the SM $SU(3)_c \times SU(2)_L \times U(1)_Y$  gauge group and does not allow proton decay. The interaction Lagrangian for $U_3$ LQ with the SM fermions is given by \cite{sakaki}
\bea
\mathcal{L}^{LQ}&=& h_{3L}^{ij}\bar{Q}_{iL} \pmb{\sigma}  \gamma^\mu L_{jL} {\bf U}_{3\mu}, \label{Lagrangian}
\eea
where $Q(L)$ is the left handed quark (lepton) doublets, $h_{3L}^{ij}$ are the LQ couplings and $\pmb{\sigma}$ represents the Pauli matrices. 
After performing the Fierz transformation and comparing  with (\ref{ham-bc}), we obtain an additional Wilson coefficient as
\bea
&&C_{V_1}^l=-\frac{1}{2\sqrt{2}G_F V_{cb}}\sum_{k=1}^3 V_{k3} \frac{h_{3L}^{2l}{h_{3L}^{k3}}^*}{M^2_{U_3^{2/3}}}. \label{CV1} 
\eea
Similarly the comparison of Eqn. (\ref{Lagrangian}) after Fierz transformation, with the SM effective Hamiltonian (\ref{ham-bs}), one can obtain  new Wilson  coefficients  $C_{9,10}^{NP}$ 
\bea
  C_{9}^{ NP} &=& -C_{10}^{ NP}  = \frac{\pi}{\sqrt{2} G_F V_{tb}V_{ts}^* \alpha} 
 \frac{h_{3L}^{ni}{h_{3L}^{mj}}^*}{M^2_{U_3^{2/3}}}\,. \label{C9-bs} 
\eea
\section{Constraint on leptoquark couplings and numerical analysis}
After knowing the new physics contribution to the SM, we now  proceed to constrain the new parameter space by using the experimental limit on the branching ratios of  $B_s \to l^+ l^-$ and $\bar B \to X_s l^+ l^-$ processes. Including LQ model, the branching ratio of  $B_s \to l^+ l^-$ process is given by 
\bea
{\rm Br}(B_s \to l^+ l^-) = \frac{G_F^2}{16 \pi^3} \tau_{B_s} \alpha^2 f_{B_s}^2 |C_{10}^{\rm SM}|^2 M_{B_s} m_{l}^2   |V_{tb} V_{ts}^*|^2
 \sqrt{1- \frac{4 m_l^2}{M_{B_s}^2}} \Bigg | 1+ \frac{C_{10}^{NP}}{C_{10}^{SM}} \Bigg |^2.
\eea
Now comparing the theoretical value of branching ratio with the $1\sigma$ range of the  experimental data, the constraint on the real and imaginary part of the LQ couplings  are given in Table 1. 

The branching ratio of $\bar B \to X_s l l $ process in the LQ model is given by
\bea
\left (\frac{d {\rm Br}}{d s_1 }\right )_{\rm LQ}&=& B_0 \Big[\frac{16}{3} (1-s_1)^2 (1+2 s_1)[{\rm Re}(C_9^{eff} C_9^{NP *}+{\rm Re}(C_{10} C_{10}^{NP *}]\nn\\
&+& \frac{8}{3}(1-s_1)^2(1+2 s_1)\left [|C_9^{NP}|^2 +|C_{10}^{NP}|^2 \right ]
+ 32 (1-s_1)^2 ~{\rm Re}(C_7 C_{10}^{NP *})  \Big], 
\eea
where $s_1=q^2/m_b^2$ and $B_0$ can be found in \cite{mohanta}. The allowed region of corresponding LQ couplings  which are compatible with the $1\sigma$
range of the experimental result in low $q^2$ are given in Table 1. 

Now using the constrained LQ couplings from Table 1 and Eqns. (\ref{CV1}, \ref{C9-bs}), we compute the bound on new Wilson coefficients $C_{V_1}^l$, $C_{9,10}^{NP}$.  In Table II, we present the predicted values of branching ratios of $B \to D^{(*)} l \nu_l$  processes  in both the SM and  $U_3$ LQ model. In Fig. 1, we show the $q^2$ variation of LNU parameters, $R_D (q^2)$ (left panel) and $R_{D^*}(q^2)$ (right panel). The plots for $R_K(q^2)$ (left panel) and $R_{K^{*}}(q^2)$ (right panel) parameters in low $q^2$ region  are shown in Fig. 2. The corresponding numerical values of LNU parameters  are given in Table 2.
\begin{table}[h]
\caption{Constraints  on the real and imaginary parts of the leptoquark couplings (for $M_{LQ}=1$ TeV) from $B_s \to l^+ l^-$ and $\bar B \to X_s ll$ processes \cite{mohanta}}
\begin{center}
\begin{tabular}{| c | c | c | c |}
\hline
Deacy process ~&~Leptoquark Couplings~ &~ Real part~ & ~Imaginary Part~  \\
\hline
&$h_{3L}^{21} {h_{3L}^{{31}^*}}$ & $-13.0 \to 13.0$ & $-13 \to 13$ \\
$B_s \to l^+ l^-$ & $h_{3L}^{22} {h_{3L}^{{32}^*}}$   &   $-0.016 \to 0.0$ &$-0.008 \to 0.008$  \\
&$h_{3L}^{23} {h_{3L}^{{33}^*}}$   &   $-0.4 \to 0.4$ &$-0.4 \to 0.4$  \\
   \hline
$\bar B \to X_s l^+ l^-$ &$h_{3L}^{21} {h_{3L}^{{31}^*}}$ & $-0.01 \to 0.01$ & $-0.01 \to 0.01$ \\
& $h_{3L}^{22} {h_{3L}^{{32}^*}}$   &   $-0.008 \to 0.008$ &$-0.008 \to 0.008$  \\
 \hline
\end{tabular}
\end{center}
\end{table}
\begin{figure}[h]
\centering
\includegraphics[scale=0.37]{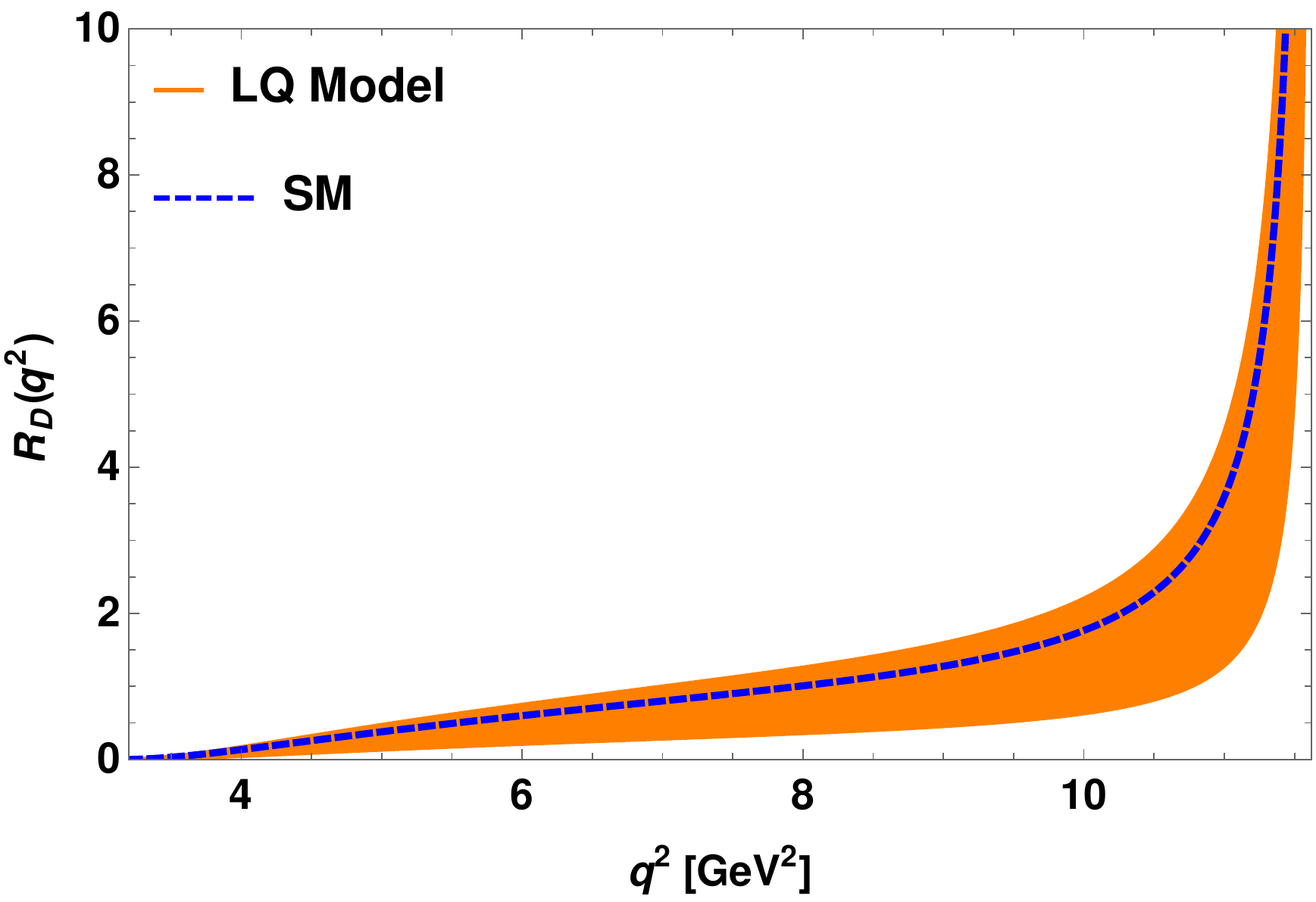}
\quad
\includegraphics[scale=0.37]{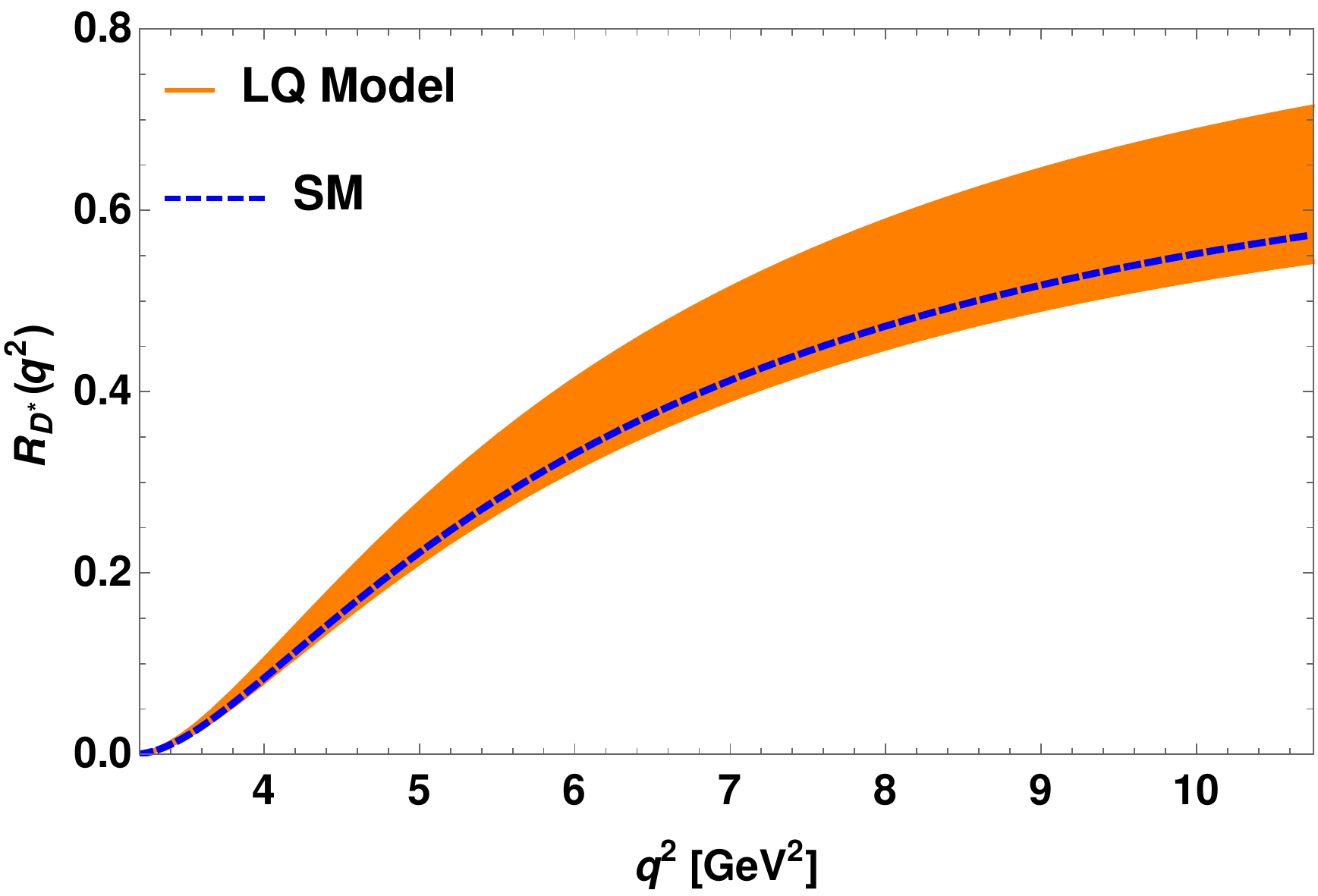}
\caption{The  $q^2$ variation of  $R_D (q^2)$ (left panel) and $R_{D^*}(q^2)$ (right panel) parameters in  LQ model.}
\end{figure}
\begin{table}[h]
\caption{The predicted values of branching ratios of $\bar B \to D^{(*)} l \bar \nu_l$  processes and the $R_{D^{(*)}} $,  $R_{K^{(*)}}$ anomalies  in  vector leptoquark model \cite{mohanta, mohanta2}}
\begin{center}
\begin{tabular}{| c | c | c | }
\hline
~Observables~ & ~SM Predictions ~ & ~Values in LQ Model~  \\
\hline
Br$\left(\bar{B} \to D l \bar{\nu}_l \right)$ & $(2.18 \pm 0.13) \times 10^{-2}$  & $(2.13-2.25) \times 10^{-2}$ \\

Br$\left(\bar{B} \to D \tau \bar{\nu}_l \right)$ & $(6.75\pm 0.08) \times 10^{-3}$  & $(2.48-8.2) \times 10^{-3}$ \\

Br$\left(\bar{B} \to D^* l \bar{\nu}_l \right)$ & $(5.18 \pm 0.31) \times 10^{-2}$ & $(5.04-5.32) \times 10^{-2}$\\

Br$\left(\bar{B} \to D^* \tau \bar{\nu}_l \right)$ & $(1.33\pm 0.14) \times 10^{-2}$  & $(1.3-1.6) \times 10^{-2}$ \\

$R_D$ & $0.31$ & $0.11-0.386$   \\

$R_{D^*}$ & $0.26$ & $0.243-0.32$ \\

${R_{K}}_{q^2 \in[1,6]}$ & $1.006$ & $0.75-1.006$ \\


${R_{K^*}}_{q^2 \in[1,6]}$ & $0.996$ & $0.725-0.996$ \\


 \hline
\end{tabular}
\end{center}
\end{table}
\begin{figure}[h]
\centering
\includegraphics[scale=0.54]{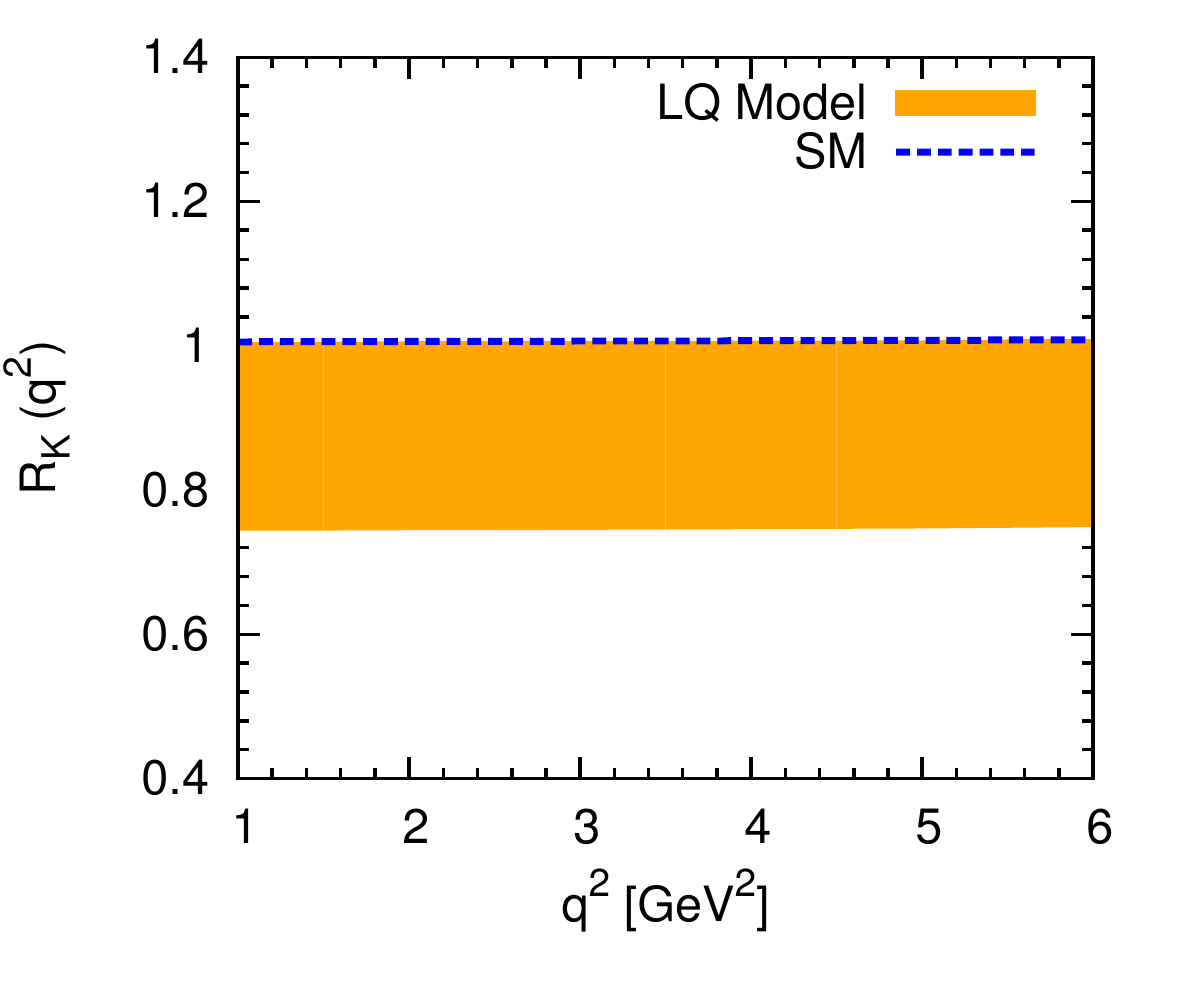}
\quad
\includegraphics[scale=0.54]{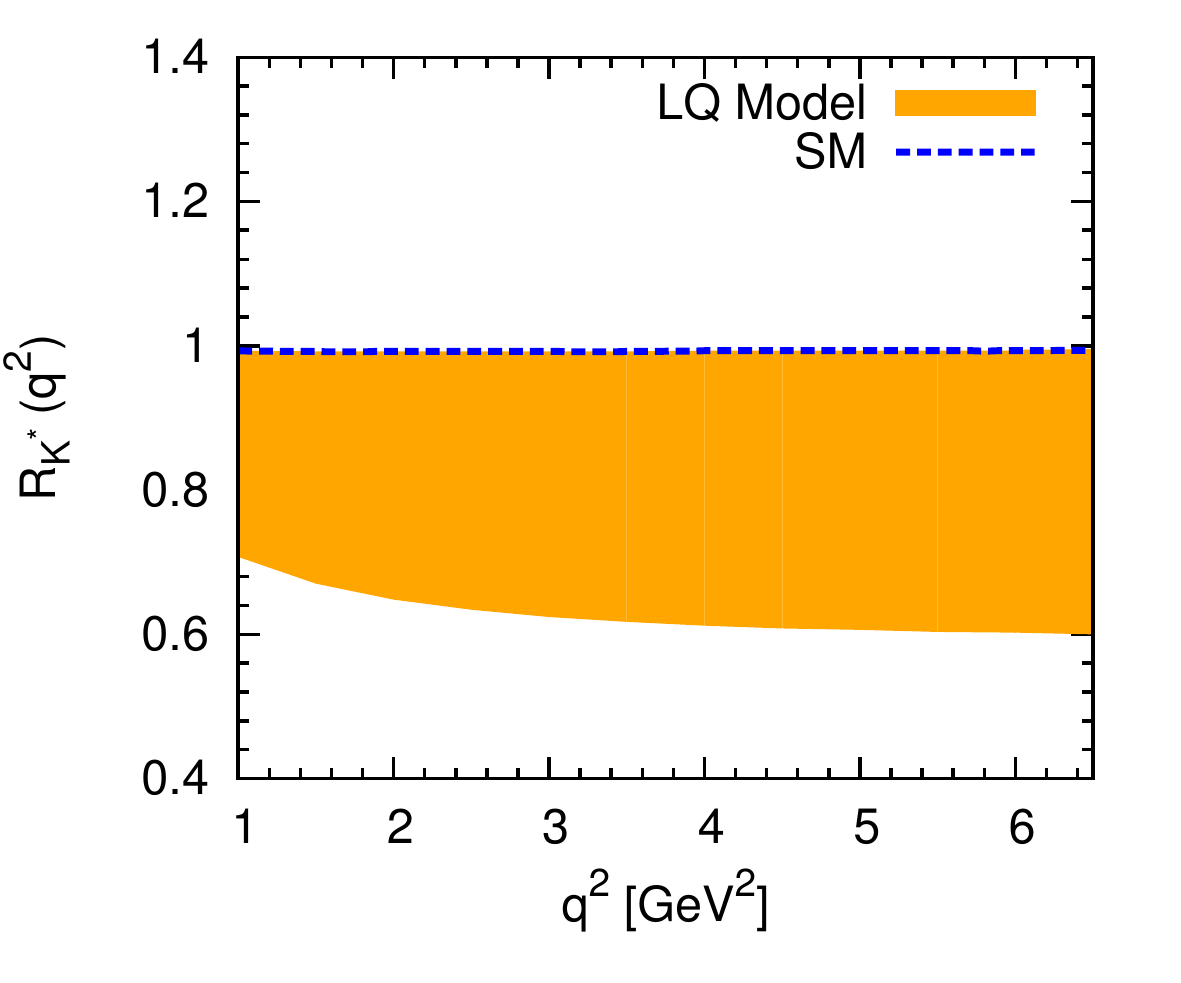}
\caption{The plot for  $R_K(q^2)$  (left panel) and $R_{K^*}(q^2)$   (right panel)  in the low $q^2$ region in  LQ model.}
\end{figure}
\section{Conclusion}
In this work we have studied the rare semileptonic $\bar B \to D^{(*)} l \bar \nu_l$ decays and the lepton nonuniversalty parameters such as $R_{D^{(*)}}$ and  $R_{K^{(*)}}$ in the $U_3(3,3,2/3)$ vector LQ model. The LQ parameter space is constrained by using the branching ratios of rare $B_s \to l^+ l^-$ and $\bar B \to X_s l^+ l^-$ decay processes. We computed the branching ratios and lepton nonuniversality in  $\bar B \to D^{(*)} l \bar \nu_l$ processes  in the $U_3 $ vector LQ model. The anomaly in $R_K$ observable is also studied. We found that the observed anomalies can be accommodated in this model.


\begin{thebibliography}{99}
\bibitem{BaBar}
BaBar Collaboration, J. Lees et al., Phys. Rev. Lett. \textbf{109}, 101802 (2012); BaBar Collaboration, J. Lees et al., Phys. Rev. D \textbf{88}, 072012 (2013).
\bibitem{Belle}
Belle Collaboration, M. Huschle et al., Phys. Rev. D \textbf{92}, 072014 (2015); Belle Collaboration, A. Abdesselam et al., [arXiv:1603.06711].

\bibitem{hfag} 
Heavy Flavour Averaging Group, 
http://www.slac.stanford.edu/xorg/hfag/semi/winter16/\\
winter16\_dtaunu.html.
\bibitem{RD-SM}
H. Na et al., Phys. Rev. D \textbf{92}, 054410 (2015).
\bibitem{LHCb}
R. Aaij et al. [LHCb Collaboration], Phys. Rev. Lett. \textbf{115}, 111803 (2015) Addendum: Phys.
Rev. Lett. \textbf{115}, 159901 (2015).
\bibitem{RD*-SM}
S. Fajfer, J. F. Kamenik, and I. Nisandzic, Phys. Rev. D \textbf{85}, 094025 (2012).
\bibitem{LHCb-RK}
R. Aaij et al., [LHCb Collaboration], Phys. Rev. Lett. \textbf{113}, 151601 (2014).

\bibitem{Hiller}
C. Bobeth, G. Hiller, G. Piranishvili, JHEP \textbf{12}, 040 (2007).


\bibitem{sakaki}
Y. Sakaki, R. Watanabe, M. Tanaka and A. Tayduganov, Phys. Rev. D \textbf{88}, 094012 (2013).

\bibitem{mohanta}
S. Sahoo, R. Mohanta and A. Giri, [arXiv:1609.04367].
\bibitem{mohanta2}
A. Giri, R. Mohanta and S. Sahoo,  J.Phys.Conf.Ser. \textbf{770}, 012031 (2016).

\end{thebibliography}
\end{document}